\documentclass[a4paper,man,natbib]{apa6}

\usepackage[english]{babel}
\usepackage[utf8x]{inputenc}
\usepackage{amsmath}
\usepackage{graphicx}
\usepackage[colorinlistoftodos]{todonotes}
\usepackage{url}

\title{Intention to explore the role of discretization in the emergence of self-organization in certain approximations of continuous cellular automata and other complex dynamic systems}
\shorttitle{Intention to explore discretization in self-organization}
\author{Q. Tyrell Davis}
\affiliation{Computer Science Department, The University of Vermont \\
Burlington, VT 0540}
\abstract{John H. Conway's Game of Life, as well as cellular automata in the larger family of Life-like CA, are discrete: the cells have a binary state space and the birth and survival transition rules are 9-bits apiece. Inspired by Life, several projects have developed continuously-valued cellular automata frameworks in 2 dimensions. These CA systems are necessarily {\it imperfect approximations} of ideal continuous systems when they are implemented in a digital computer, and this inevitably leads to discretization errors. As we know since at least the time of Poincairé's work on the three-body problem, arbitrarily small errors in a complex dynamic system can lead to substantial behavioral deviation over time. 

This outline of intent is based on observations in cellular automata that, in certain cases, errors are not only well-tolerated, but are essential for self-organization. 

This manuscript describes a set of experiments to investigate the importance of discretization in approximating continuously valued dynamic systems, how and when discretization is required for self-organization to occur, and the relevance the results might have with respect to emergent agency and analogies for consciousness. }

\begin{document}
\maketitle
\section{Preface}
This intention manuscript fulfills a similar role to a Stage 1 article (or paper proposal) in the peer review process known as registered reports: it lays out background, preliminary results, and a plan for experiments with discussion of what experimental results will mean if they turn out one way or another. Registered reports were introduced as a peer-review strategy for addressing replication issues, post-hoc hypothesizing, and multiple types of bias while reducing perverse career incentives in academic research\footnote{{\it e.g.} https://web.archive.org/web/20220707161327/https://neurochambers.blogspot.com/2012/10/changing-culture-of-scientific.html}. See \citep{chambers2022} for a review of registered reports and their impact on publications in the field of psychology and beyond. 

Many of the most interesting contributions to the field of complex systems, and cellular automata in particular, exist outside peer review, the product of tinkerers, engineers, artists, and anyone else who wants to quickly and informally declare ``look what I built.'' Those sources are referenced throughout in footnotes. 

\section{Introduction}
\label{intro}

\begin{figure}[htb]                                       
\begin{center}                                              
  \includegraphics[width=0.8\textwidth]{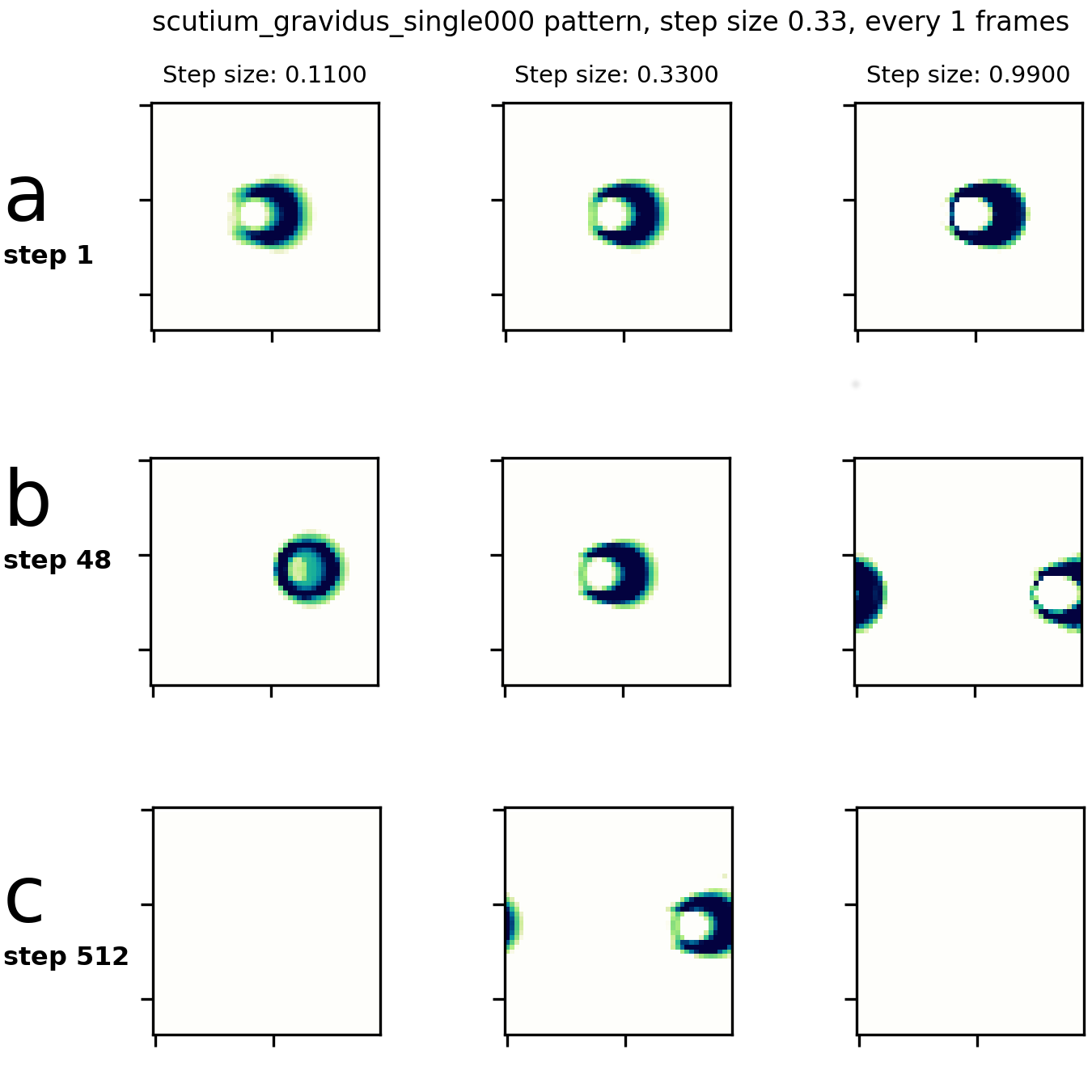}
    \caption{A small glider in the {\it Scutium gravidus} rules in Lenia becomes unstable for step sizes that are too small or too high. {\bfseries a)} step 1; {\bfseries b)} step 48 {\bfseries c)} step 512. }
  \label{fig:scutium_dt}
\end{center}        
\end{figure}

\begin{figure}[htb]                                       
\begin{center}                                              
  \includegraphics[width=0.8\textwidth]{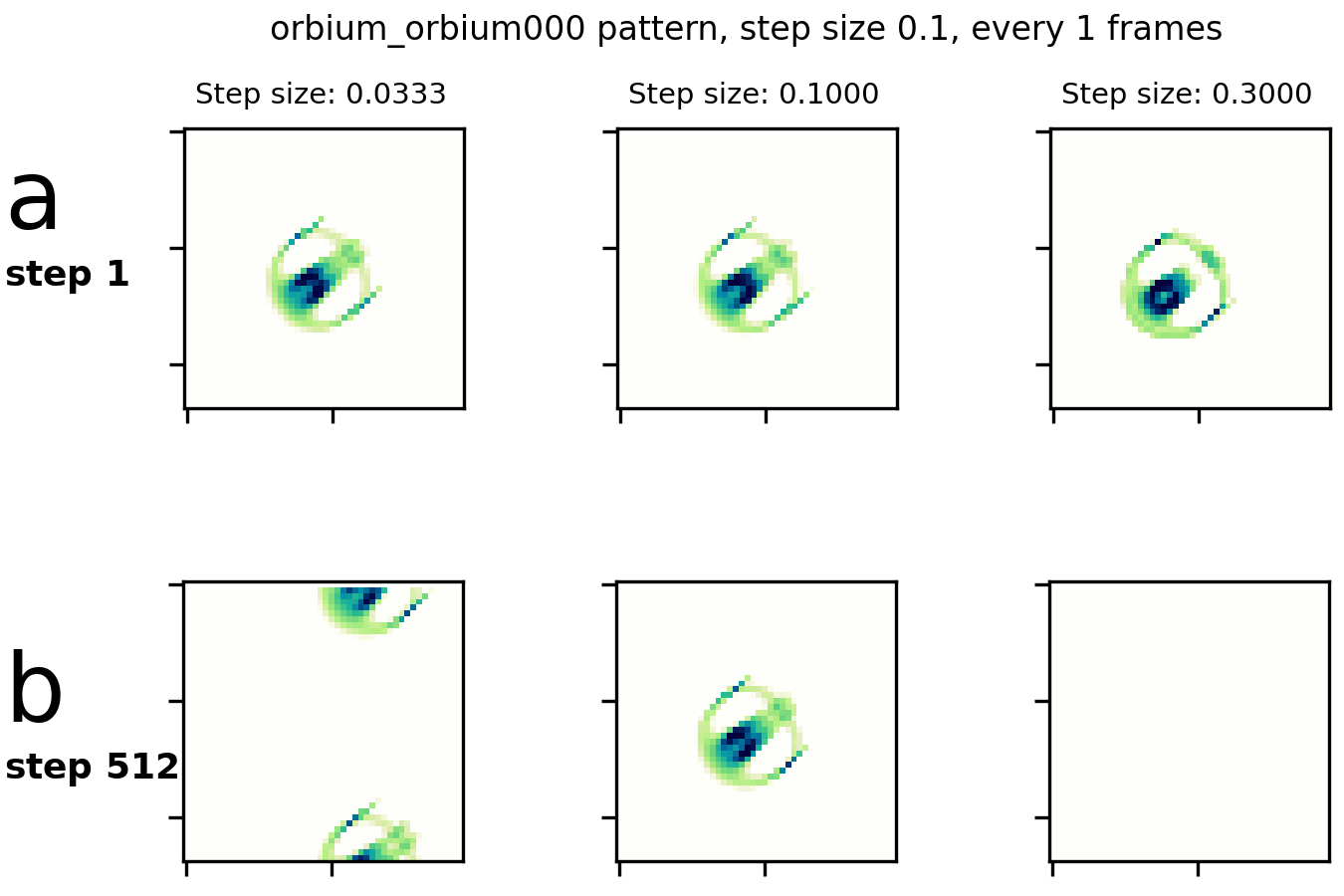}
    \caption{The original {\it Orbium} glider in Lenia with Gaussian growth function parameters $\mu_G=0.15, \sigma_G=0.015$ and Gaussian neighborhood kernel parameters $\mu_K=0.5, \sigma_K=0.15$. Only the too-large step size destabilized the glider. {\bfseries a)} step 1; {\bfseries b)} step 512. }
  \label{fig:orbium_dt}
\end{center}        
\end{figure}

The emergence of self-organized individuality and activity in cellular automata (we will refer to both the plural and singular as CA) is well-represented in the story of the reflex glider from Conway's Game of Life \citep{berlekamp2004}. While manually simulating the rules on a Go board, Richard Guy reportedly exclaimed ``Come over here, there's a piece that's walking!'' He had discovered the simple reflex glider. The reflex glider in Life is an example of the emergence of self-organized entities, a pattern with some ability to maintain its pattern while moving and interacting with its surroundings.  

Life's glider would become an essential building block of computational universality in Conway's Game. Gliders, a term we will use in this article to include larger mobile patterns known as spaceships as well as minimal movers like the reflex glider, fulfill computational roles of information transfer and information processing. They feature prominently in Life computers, including a universal Turing machine described in \citep{rendell2011}, the OTCA metapixel by Brice Due\footnote{\url{https://otcametapixel.blogspot.com/}}, and the p1 megacell\footnote{\url{https://conwaylife.com/wiki/P1_megacell}} and 0E0P metacell\footnote{\url{https://cp4space.hatsya.com/2018/11/12/fully-self-directed-replication}}, both by Adam P. Goucher. 

Gliders have also been of interest in computer implementations of continuous CA. The SmoothLife glider, described by Stephen Rafler on arXiv\footnote{\url{https://arxiv.org/abs/1111.1567}}, is the first report of a glider in a spatially continuous CA (albeit with a discrete time step) that this author is aware of. Gliders in reaction-diffusion systems, such as the Gray-Scott system with U-Skate World parameters, were apparently discovered earlier, in 2009\footnote{\url{http://www.mrob.com/pub/comp/xmorphia/catalog.html}}. A close cousin to the SmoothLife glider was discovered in the Lenia continuous CA framework described in \citep{chan2019}, under {\it Scutium gravidus} rules, alongside hundreds of self-organizing patterns described in a comprehensive taxonomy (with an online catalog\footnote{\url{https://chakazul.github.io/Lenia/JavaScript/Lenia.html}}). Gliders can also be generated in Multiple Neighborhood Cellular Automata\footnote{See \url{https://softologyblog.wordpress.com/2018/03/09} and \url{https://github.com/slackermanz/vulkanautomata}} (MNCA, similar to Lenia's expanded universe \citep{chan2020}). CAPOW, an earlier continuous CA framework developed in the 1990s \cite{rucker2003}\footnote{see also  \url{https://github.com/rudyrucker/capow}, \url{https://www.rudyrucker.com/capow}, and \url{https://arxiv.org/abs/1111.1567}}, apparently did not lead to the discovery of gliders but did produce active Turing patterns. Likewise, an even earlier continuous treatment of cellular automata based on field computation did not yield or aim to find gliders\footnote{Bruce MacLennan 1990 \url{https://library.eecs.utk.edu/storage/564phpwKMVQwut-cs-90-121.pdf}}. 

%See also \url{https://softologyblog.wordpress.com/2018/03/09} and \url{https://github.com/slackermanz/vulkanautomata} 

Primordia\footnote{\url{https://github.com/Chakazul/Primordia}} a precursor to Lenia with multiple states and MergeLife with more than 16 million \citep{heaton2019} (in 24-bit RGB format\footnote{{\it i.e.} $2^{24} = 16,777,216$ possible states}), are intermediate between Life-like CA and continuous CA, and both have rules that give rise to gliders. % other examples

Most continuous CA, especially continuous spatial automata (MacLennan), CAPOW \citep{rucker2003}, SmoothLife (Rafler), and Lenia \citep{chan2019}, were developed as mathematically ideal continuous systems with necessarily digitized computational implementations. 

For example, to approximate continuous Lenia described as

\begin{equation}
    A^{t+dt} = A^t + dt G(K*A^t) 
\end{equation}

where $A^t$ is the grid cell states at time t, $G(\cdot)$ is the growth (or transition) function, $K$ is the neighborhood kernel and $*$ represent 2D convolution, we estimate the continuous rate of change as the change in an arbitrarily small time step $\Delta t$ 

\begin{equation}
    A^{t+\Delta t} = \lim\limits_{\Delta t \to 0} A^t + \Delta t G(K*A^t) 
\end{equation} 

In other words we use Euler's method to estimate continuous CA dynamics defined by the differential equation

\begin{equation}
    \frac{dA}{dt} =  G(K*A^t) 
\end{equation}

With the same construction as Euler's method, we would generally expect that the smaller we make the step size $\Delta t$, the more accurate the simulation. However, previous work by this author \citep{davis2022a} showed that in some cases gliders lose their ability to self-organize for step sizes that are too small. In the Preliminary Results Section we will see that, for some patterns, more closely approximating a continuous ideal by using a larger, smoother neighborhood kernel or by using higher precision floating point data types can also cause a pattern to lose its self-organized identity. 

\section{Hypotheses}
\label{sec:hypotheses}

\begin{enumerate}
    \item In computational implementations approximating continuous complex dynamic systems, approximation parameters (spatial and temporal resolution, numerical precision, etc.) can be as important to the ability of patterns to self-organize as are the nominal parameters of the system.
    \item The importance of approximation parameters to self-organization is not limited to a particular continuous CA framework, or even to continuous CA, but also can be observed in other discrete approximations of other complex systems that support self-organizing patterns.
    \item The sensitivity of a self-organizing pattern, such as a glider, to approximation parameters depends on the context in which it is selected from. 
\end{enumerate}

To be clear, when I say that self-organization is sensitive to approximation parameters, I mean that this sensitivity is {\it double-sided}. That is, for some patterns the systematic simulation errors that accumulate as a result of approximation coarseness are essential for self-organization. As we would expect, failures also occur when approximations are too coarse.

\section{Preliminary Results}
\label{sec:results}

\begin{figure}[htb]                                       
\begin{center}                                              
  \includegraphics[width=0.75\textwidth]{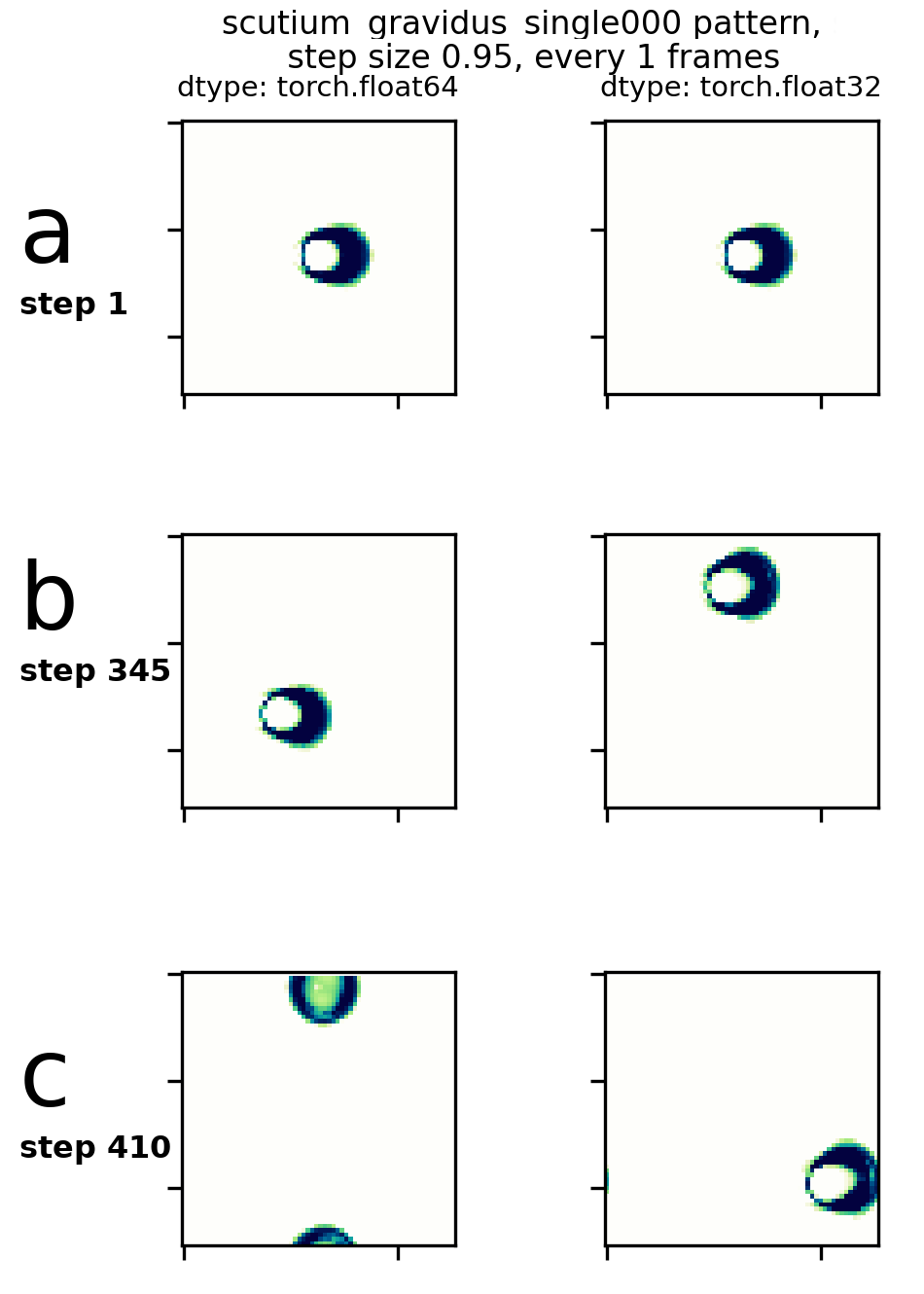}
    \caption{{\it Scutium gravidus} glider simulated with {\tt torch.float64} and {\tt torch.float32} precision. {\bfseries a)} step 1; {\bfseries b)} step 345 {\bfseries c)} step 410. }
  \label{fig:scutium_dtype}
\end{center}        
\end{figure} 

\begin{figure}[htb]                                       
\begin{center}                                              
  \includegraphics[width=0.75\textwidth]{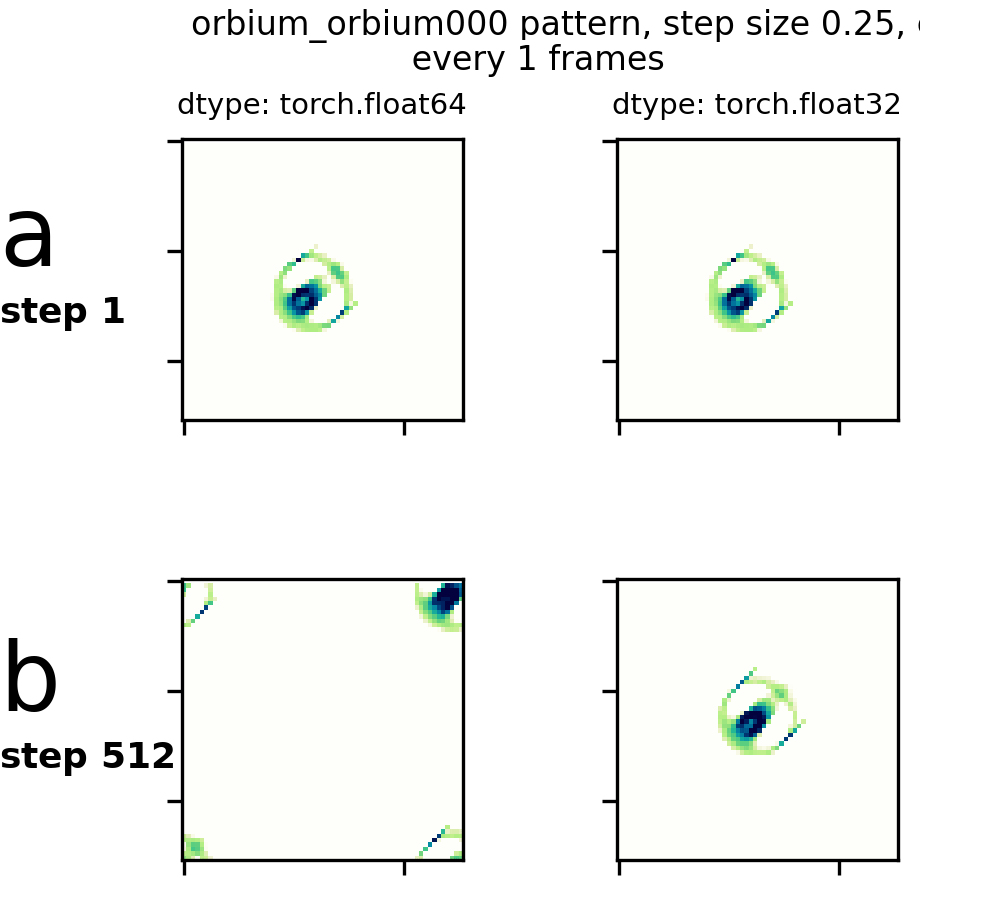}
    \caption{{\it Orbium} glider simulated with {\tt torch.float64} and {\tt torch.float32} precision. {\bfseries a)} step 1; {\bfseries b)} step 512.}
  \label{fig:orbium_dtype}
\end{center}        
\end{figure} 

\begin{figure}[htb]                                       
\begin{center}                                              
  \includegraphics[width=0.75\textwidth]{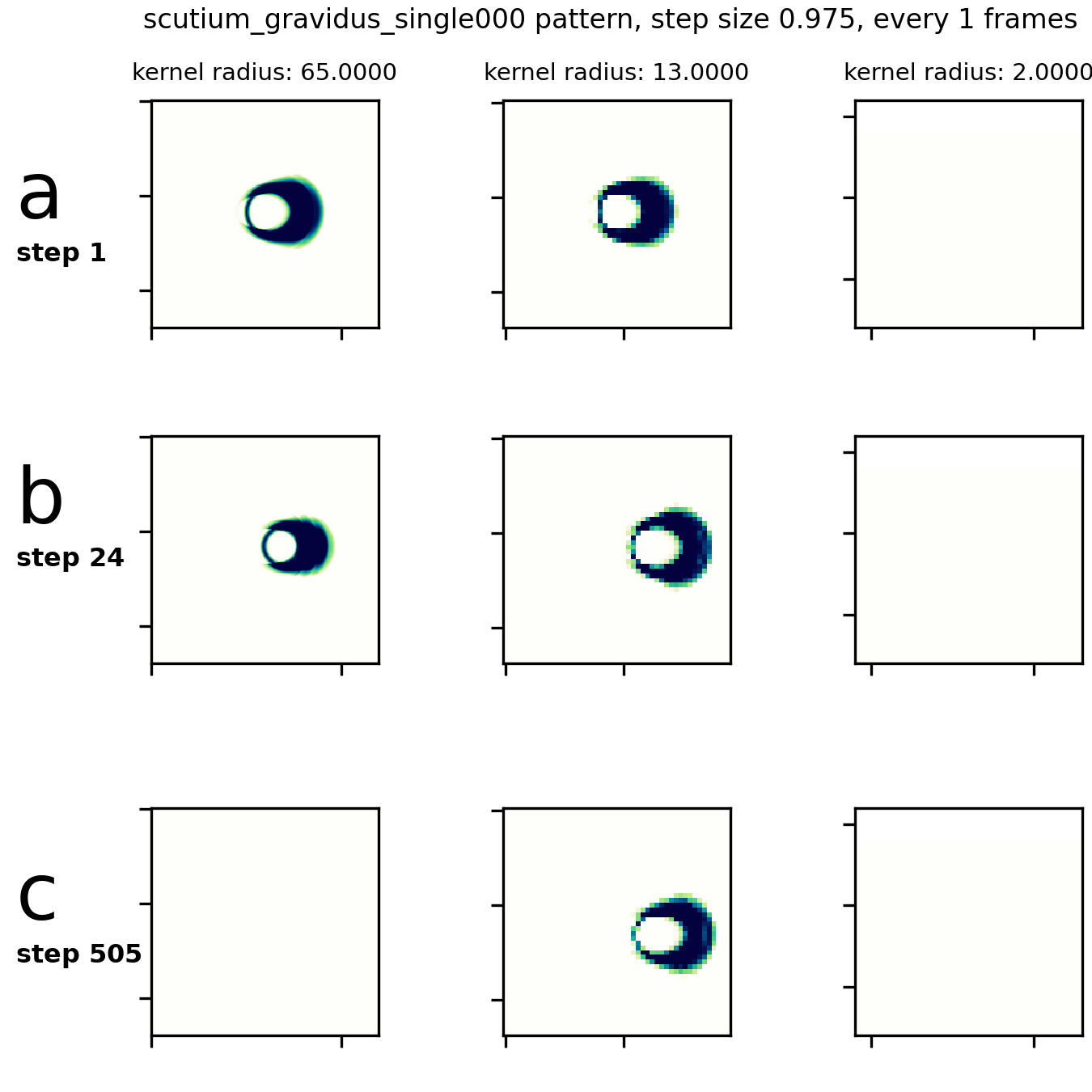}
    \caption{{\it Scutium gravidus} glider simulated with kernel radius 65, 13, and 2. Larger kernel size corresponds to better spatial resolution (smaller $\Delta_{x,y}$ {\bfseries a)} step 1; {\bfseries b)} step 24 {\bfseries c)} step 505. }
  \label{fig:scutium_kernel}
\end{center}        
\end{figure} 

\begin{figure}[htb]                                       
\begin{center}                                              
  \includegraphics[width=0.75\textwidth]{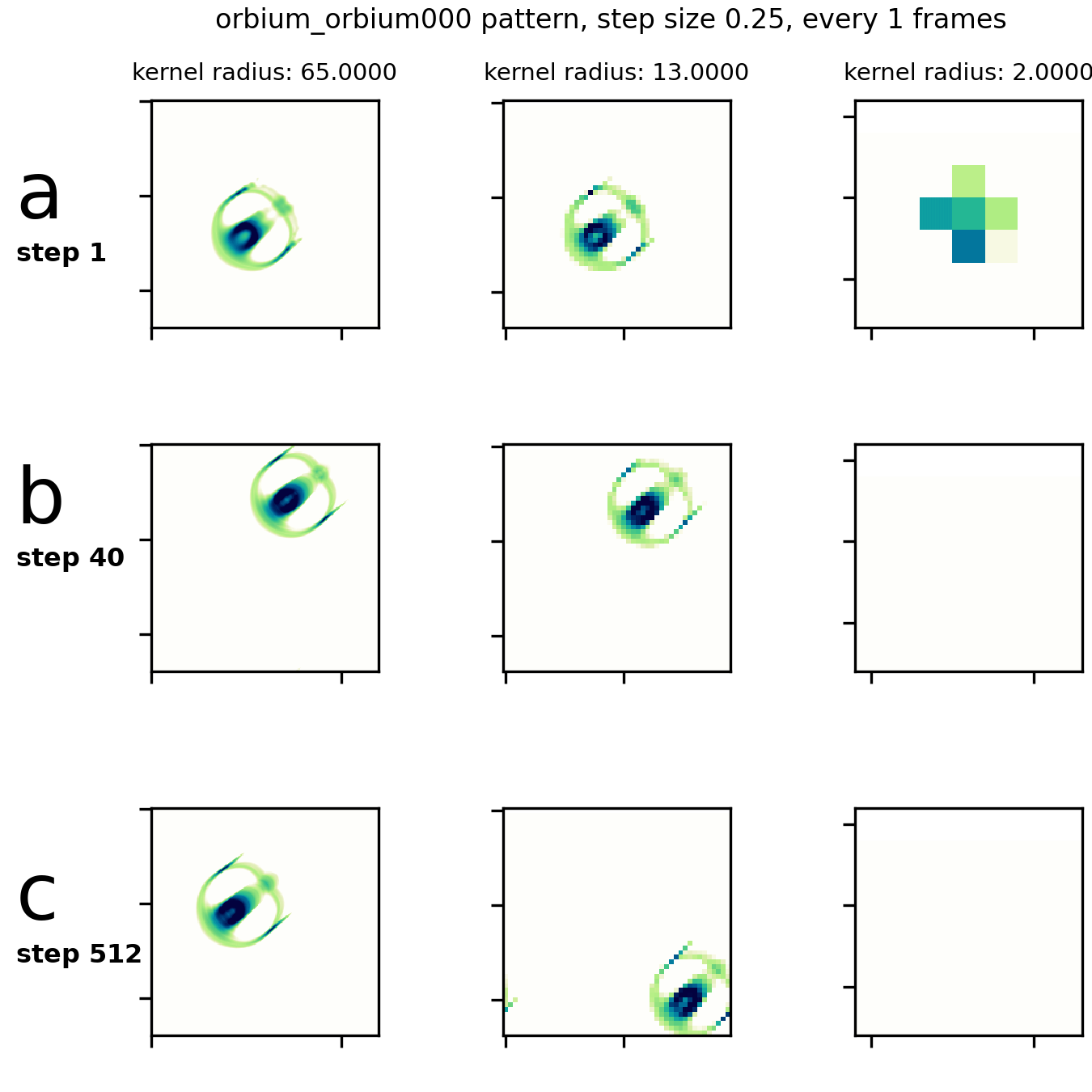}
     \caption{{\it Orbium} glider simulated with kernel radius 65, 13, and 2. Larger kernel size corresponds to better spatial resolution (smaller $\Delta_{x,y}$ {\bfseries a)} step 1; {\bfseries b)} step 40 {\bfseries c)} step 512. }
  \label{fig:orbium_kernel}
\end{center}        
\end{figure}

The figures referred to in this section show time-lapse snapshots of a small glider pattern in {\it Scutium gravidus} and the {\it Orbium} glider, both are instances of CA in the Lenia framework \citep{chan2019}. These figures show that increasingly accurate approximations of a mathematical ideal continuous CA system are well-tolerated by the {\it Orbium} pattern, but lead to instability for {\it Scutium}. These figures show only a few frames, full animations are available at \url{https://rivesunder.github.io/DisContinuous}, as well as links to code for replicating the simulations they represent. 

Preliminary results consist of observations of {\it Orbium} and {\it Scutium} gliders while varying temporal step size $\Delta t$, spatial resolution $\Delta_{x,y}$ (via kernel size), and numerical precision (single and double precision with {\tt torch.float32} and {\tt torch.float64} in deep learning library PyTorch \citep{pytorch}\footnote{https://github.com/pytorch/pytorch}, respectively). 

\subsection{Temporal Resolution}

In keeping with previously described observations \citep{davis2022a}, step sizes that are either too small or too large can cause a Lenia glider to lose its ability to pull itself together, or self-organize.

In Figure \ref{fig:scutium_dt} a {\it Scutium gravidus} glider is unstable at a moderately small step size of 0.11 and a large step size 0.99, while the glider remains stable over 512 steps with a $\Delta t$ of 0.33. {\it Orbium} on the other hand is stable at much smaller step sizes but becomes unstable at large step sizes. Figure \ref{fig:orbium_dt} includes time-lapse frames from an {\it Orbium} glider remaining stable for 512 steps at step sizes of 0.033 and 0.1, although it is unstable at a larger step size of 0.3.

\subsection{Numerical Precision}

A simulation with higher-precision floating point numbers led to instability for a {\it Scutium gravidus} glider (Figure \ref{fig:scutium_dtype}). The glider simulated in PyTorch's {\tt torch.float64} data type persists for hundreds of steps (although it deviates from the path taken under default {\tt torch.float32} precision), but eventually vanishes after 410 steps. {\it Orbium} tolerates double precision simulation, gliders persist for 512 steps under single and double precision (although paths differ). 

\subsection{Spatial Resolution}

A larger neighborhood kernel corresponds to a smaller spatial resolution and smoother simulation. As in the case of numerical precision and temporal resolution, the {\it Scutium gravidus} glider becomes unstable under a closer approximation of the mathematical ideal of a truly continuous CA (Figure \ref{fig:scutium_kernel}). The {\it gravidus} glider fails to persist with a neighborhood kernel with radius 65. The {\it Orbium} glider again tolerates a smoother simulation, persisting for 512 steps with a kernel radius of either 65 or 13 (Figure \ref{fig:orbium_kernel}).

\newpage

\section{Experiment Plan}
\label{sec:plan}

Previously published results \citep{davis2022a} and findings described in the Preliminary Results Section agree with Hypothesis 1 that approximation parameters are in some cases essential to self-organization. Self-organizational and behavioral sensitivity to smaller step sizes, nominally amounting to a closer approximation to the mathematical ideal of a continuous CA, was observed in both the Lenia CA framework \citep{chan2019} and in a slightly more expressive variant of Lenia, glaberish \citep{davis2022b}. Testing Hypothesis 2 means extending experimental observations to other complex systems, namely reaction-diffusion systems (such as the Gray-Scott model \citep{pearson1993, lee1993}) and neural CA, {\it e.g.} \citep{wulff1992, li2002, mordvintsev2020, niklasson2021}.

To generate novel glider pattern/rule set pairs, I will use modified versions of the evolutionary approach described in \citep{davis2022c}. In particular the two-step evolutionary process (separate stages for rule and pattern evolution) will be combined in a single co-evolutionary selection process. The co-evolution algorithm will be adapted as needed to support arbitrary complex systems such as the Gray-Scott reaction diffusion model and neural CA. Gliders from all 3 systems will then be simulated across a range of approximation parameters, producing a multidimensional map of approximation support for each pattern. If gliders from the Gray-Scott and neural CA systems also show a loss of self-organization when approximation parameters simulate the ideal continuous system with too much accuracy, this will indicate a failure to disprove Hypothesis 2. 

The combined co-evolution method will also include tools for setting approximation parameters during evolution, varying them according to some rule (such as maintaining a chosen error threshold), and evolving under a range of approximation parameters sampled from a distribution. This latter capability will provide evidence to disprove or fail to disprove Hypothesis 3: glider patterns evolved under simulation with a wider distribution of a given approximation parameter should be less sensitive to specific values of those parameters, unless the hypothesis is wrong. 

Hypotheses 1 and 2 will find support or negation in the simple presence of patterns  that lose their ability to self-organize outside of particular ranges of approximation parameters. Hypothesis 3, however, suggests a dependence of the sensitivity of self-organization to approximation parameters on the range of those parameters encountered during evolution. Therefore, Hypothesis 3 will require comparison of populations evolved under set or variable approximation parameters with an appropriate statistical test, {\it e.g.} Student's one-tailed t-test, with at least 3 population comparisons in total (one for each of the 3 systems under study). 

\subsection{Outline of experiments}

\begin{enumerate}
    \item Map approximation parameter support for previously described self-organizing glider patterns in simulations of continuous CA: at least one pattern each from the 3 Lenia classes catalogued in the Lenia web demo\footnote{\url{https://chakazul.github.io/Lenia/JavaScript/Lenia.html} classes include Exokernel, Mesokernel, Endokernel}, at least 3 glider patterns evolved in the Glaberish framework \citep{davis2022b}, and at least 3 known reaction-diffusion gliders from the U-Skate World catalog referenced earlier.
    \item Co-evolve simulation rules and glider patterns in both neural CA and reaction diffusion systems. A subsection ($\approx$ half) of these rule-pattern pairs will be evolved with variable approximation parameters, {\it e.g.} sampled from a Gaussian distribution.
    \item Map approximation parameter support for novel, evolved CA gliders (including neural CA), and for gliders from the reaction-diffusion model.
    \item Compare the variance between glider/rule pairs evolved with and without variable approximation parameters to determine presence or absence of a statistical difference. 
\end{enumerate}

\section{Relevance}
\label{sec:discussion}

\subsection{A surprise in the context of previous work}
\label{sec:perplexity}

The first taste of the relevance of this work to the field of cellular automata in particular and complexity in general is one of perplexity. Several different authors have discussed the role of continuity in computational CA systems approaching continuous simulation, but none have remarked on the phenomenon of essential coarseness for self-organizing behavior \citep{chan2019, rucker2003} (and Rafler 2012,  MacLennan 1990). Instead the views expressed in those works indicate the intuitive expectation that less accurate simulation may destroy active patterns and self-organizing gliders, but not more accurate simulation.

The observations described in the Preliminary Results Section and \citep{davis2022a} run counter to the expectation of continuous CA as ideal mathematical systems, at least with regard to the interesting behavior of self-organizing gliders when we implement these systems in a computer. 

In \citep{rucker2003}, the author experimented with higher-precision floating point numbers (doubles) but found the change to be inconsequential to observed behavior, leading them to use lower-precision single floats instead. Lower-precision numbers ({\it i.e.} 10-bits), on the other hand, led to vanishing patterns in CAPOW. Rafler described the use of anti-aliasing apodization of neighborhood kernels, and the use of sigmoidal transition functions to smooth out otherwise abrupt edges. While the SmoothLife glider was discovered in simulations with a large discrete time step (equivalent to a $\Delta t$ of 1.0), the role of discretization in glider support was not discussed. In Lenia, we also find a close cousin to the discrete time-step SmoothLife glider: a small glider in the {\it Scutium gravidus} rule set (one of the glider investigated in the Preliminary Results Section), stable in a range of relatively large time steps, but not small ones.

In \citep{chan2019}, Bert Chan discussed discretization as a trade-off in accuracy and computational requirements, with increasing kernel sizes ({\it i.e.} decreasing $\Delta_{x,y}$), increasing numerical precision, and decreasing $\Delta t$ step size all more closely approaching a mathematical ideal as discretization tends to 0. Chan focused on the {\it Orbium} glider when considering discretization. As evident in the Preliminary Results Section, {\it Orbium} does meet intuitive expectations as discretization tends to 0, but that characteristic is not shared with other pattern/rule pairs in Lenia CA in general.

\subsection{Potential relevance to agency under consistent physics}
\label{sec:agency}

An understanding of discretization as it relates to self-organization in complex systems may help facilitate the development of emergent patterns, like gliders, with increased agency. The ability of the Life reflex glider to interact with its environment, its so-called cognitive domain, is limited; there are only a handful of perturbations that the glider can survive \citep{beer2014}. Glider patterns in Lenia, however, have exhibited a range of survivable interactions with environmental objects when they are optimized for the purpose with gradient descent as described by Hamon {\it et al.} \footnote{Hamon, et al., ``Learning Sensorimotor Agency in Cellular Automata", 2022. \url{https://developmentalsystems.org/sensorimotor-lenia/}}. 

In this author's opinion, the emergence of agency in simulated worlds under consistent physics is probably the most promising line of research for (dis)continuous cellular automata and similar complex systems, and the greatest justification for the added complication and computational requirements of said systems. Unlike the conventional regime of reinforcement learning, where the environment and agents are wholly separate from each other, persistent patterns in CA operate under the same rules as everything else in their universe.  

\subsection{Potential links to perception and consciousness}
\label{sec:perception}

At the risk of veering into the realm of philosophy, discretization effects on self-organization in simulated complex systems may have interesting relevance to consciousness and perception, if only as a source of analogical mental tools. Whether consciousness and perception are discrete, continuous, or a mix of both, remains an unanswered question and active area of debate in the literature \citep{vanrullen2003, fekete2018}. Self-organization under the influence of systematic discretization errors could echo the role of film in providing metaphors for perception, such as the wagon wheel effect: well understood to be a result of aliasing with camera frame rate on film but that can also be experienced by human observers in continuous daylight \citep{purves1996}. 

\section{Final Remarks}

In this article I presented evidence that, in certain cases, patterns depend on what we would normally describe as systematic errors for their self-organization abilities. This runs counter to our intuitive expectation of making accuracy-computation trade-offs until reaching an acceptable level of accuracy with respect to an idealized mathematical system. The findings results described in the Preliminary Results Section and in previous work \citep{davis2022a} are consistent with the idea that approximation parameters can be as important for self-organization as nominal rule parameters. Why is this so? I suspect that the answer is a variant of the evolutionary adage ``you get what you select for.'' That is to say, glider patterns are likely to reflect the specific conditions under which they were engineered, discovered, or evolved. 

The ideas delineated in the Hypotheses Section and the outline in the Experiment Plan Section are designed to answer questions about the role of discretization in computational approximations of complex continuous systems. Namely, Hypotheses 1 and 2 will be used to test the generality of discretization on self-organization in other systems, while Hypothesis 3 is intended to probe the link between evolution (via implicit or explicit selection) and discretization as it relates to emergent self-organization.

\section*{Acknowledgments}

This work was supported by the National Science Foundation under the Emerging Frontiers in Research and Innovation (EFRI) program (EFMA-1830870).

\footnotesize
\bibliographystyle{apalike}
\bibliography{references} % replace by the name of your .bib file

\end{document}